\documentclass[final,3p,times,twocolumn]{elsarticle}
\usepackage{amssymb}
\usepackage{amsmath}
\journal{Journal of Physics and Chemistry of Solids}

\begin{document}

\begin{frontmatter}



\title{Half-Heusler TiXSn (X = Pd, Pt and Ni): electronic, vibrational, and defect properties from first-principles calculations}

\author[label1]{Mateus Corradini Lopes}
\affiliation[label1]{organization={Gleb Wataghin Institute of Physics, State University of Campinas},
%%             addressline={Rua Sérgio Buarque de Holanda, 777},             
             city={Campinas},
             postcode={13083-859},
             state={São Paulo},
             country={Brazil}}
\author[label2]{Alex Antonelli}
\affiliation[label2]{organization={Gleb Wataghin Institute of Physics and Centre for Computational Engineering \& Sciences, State University of Campinas},
             city={Campinas},
             postcode={13083-859},
             state={São Paulo},
             country={Brazil}}

%\author{}
%
%\affiliation{organization={},%Department and Organization

\begin{abstract}
The knowledge of  Half-Heusler compounds have attracted much attention as materials for thermoelectric applications. In this work, we investigate, using first-principles calculations, the electronic, vibrational, and defect properties of TiXSn (X=Ni, Pd, Pt) half-Heusler compounds. The knowledge of such properties is vital for the understanding and improvement of thermoelectric transport properties of these materials. The band gap of the three compounds increase with the atomic number of the group 10 elements, in agreement with previous findings. The electronic effective masses of the three compounds are similar, while the heavy hole effective mass of TiPtSn is larger than those of the other two materials. Our calculations of the phonon dispersion included the calculation of the LO-TO splitting indicating that TiNiSn has a stronger ionic character and polar scattering of charge carriers by optical phonons in the case of low doping. Calculation of the formation energy of Ni interstitial defect in TiNiSn is very low, in agreement with previous results. Surprisingly, for the Pd interstitial in TiPdSn the formation energy is negative, suggesting that the full-Heusler structure can be more stable than the half-Heusler one. On the other hand, the formation energy of an interstitial in TiPtSn is significantly higher, suggesting smaller effects on both electronic structure and transport properties. The formation energy of all substitutional defects investigated are substantially higher than that of interstitial ones, suggesting that they should occur in very low concentrations.
\end{abstract}

%\includegraphics[scale=0.4]{TiXSn_graphical_abstract.png}

%\item LO-TO splitting decreases with electronegativity of group 10 transition metal.
%\item Negative formation energy for interstitial Pd defects in TiPdSn.
%\item Low formation energy for Ni interstitial defects in TiNiSn.
%\item High formation energy for Pt interstitial defects.
%\item Very high formation energy for all types of substitutional defect tested in TiSnX where X = Ni, Pd and Pt.
%\item Study of physical properties of little-explored half-Heusler thermoelectric materials such as TiSnX where X = Pd and Pt.


\end{frontmatter}


\section{Introduction}
\label{sec:intro}
Half-Heusler (hH) compounds are gaining great relevance nowadays due to their potential application in thermoelectric devices such as generators and refrigerators. In addition, they also present a low aggressiveness to the environment and are earth-abundant \cite{buffon16,HERMET2016248,BERCHE201862}.

Thermoelectric properties depend closely on the presence of defects in the material, since they can determine the position of the Fermi level, the band gap, and play an important role in the transport of charge carriers. Therefore, aside from the knowledge of electronic band structure and phonon dispersion, investigating defects is vital for the design and synthesis of efficient materials. 

In particular, an increasing attention has been devoted to hH compounds TiXSn, in which one of the transition metal atoms in the compound belongs to group 10 of the periodic table: X = Ni, Pd, and Pt \cite{Aliev1990,Grykalowska2005,Hazama_Asahi_Matsubara_Takeuchi_2010,Kaur2017,BERCHE201862,ma11060868,Zheng2020,Xiong2022,Xiong2024}. 

It has been found that interstitial defects play an important role in materials such as TiNiSn \cite{Hazama_Asahi_Matsubara_Takeuchi_2010,ma11060868}. Due to the low formation energy of Ni interstitial defects, the synthesis of a defect-free material is unlikely, therefore, computing the properties of defect-free materials is not enough in order to understand the characteristics of actual materials.

We investigate the formation energies of interstitial and substitutional defects in TiXSn where X = Ni, Pd and Pt, through first-principles calculations. We have also investigated the electronic structure and phonon dispersion of these materials in order to understand their main features and establish their differences. We could observe that the interstitial defects of Pd in TiPdSn present a substantial low formation energy, which would be a relevant factor for its synthesis, since these defects significantly alter the gap of the mentioned materials. We also reproduced well-known electronic, vibrational and defect properties of TiNiSn \cite{conlinet14,andrea15} in order to validate the methodology applied to other materials.

Using density functional theory (DFT) and density functional perturbation theory (DFPT) formalisms, implemented in the Quantum Espresso package \cite{Giannozzi_2009,Giannozzi_2017}, we were able to obtain the relaxed structure of the supercells, the electronic band structure of the materials, the phonon dispersion as well as defect properties of the mentioned hH compounds.

In this work, we investigate whether the same features presented by TiNiSn with respect to the nickel interstitial defect, occur in TiPdSn and TiPtSn. Therefore, we calculated the formation energy of palladium interstitial defects in TiPdSn and platinum interstitial defects in TiPtSn in the $x \approx 0.012$ concentration. Additionally, we calculated the formation energy of substitutional defecets as it will be described in the methodology section. We also have reproduced the results for Ni replacing Ti and Sn in TiNiSn \cite{Hazama_Asahi_Matsubara_Takeuchi_2010}, as well as calculating Ti and Sn replacing Ni, in order to verify the reliability of the applied methodology.

This article is organized as follows: in section \ref{sec:materials} we present the structural aspects of the defects in these materials we study, describing in detail the models for each type of defect. In section \ref{sec:methodology} we show a theoretical overview of the work, describing all the rationale involved in the research, then in section \ref{sec:results} we provide the results and make a brief discussion of them. Finally, in section \ref{sec:conclusions} we present our conclusions.

\begin{figure} [!h]
	\centering
	\includegraphics[scale=0.3]{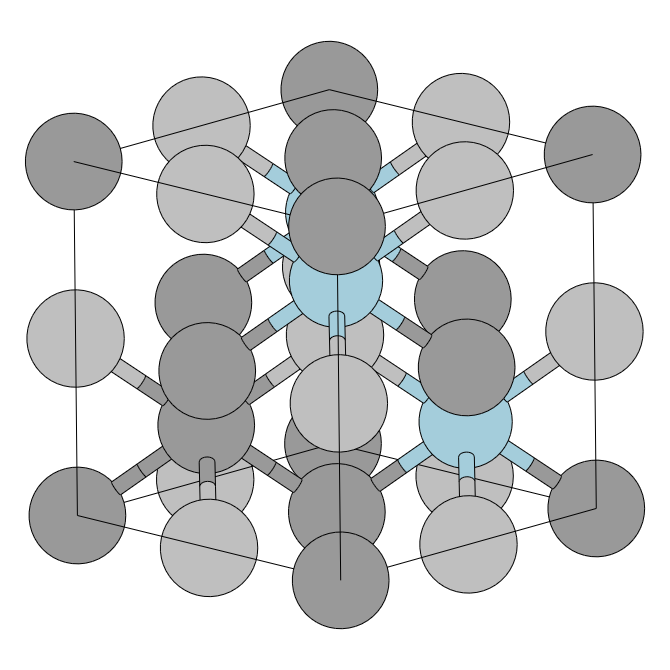}
	\caption{TiNiSn hH conventional cell. Light blue refers to Ni atoms, light gray to Sn atoms and dark gray to Ti atoms.}
	\label{fig:hh}
\end{figure}

\section{Materials}\label{sec:materials}
 
hH materials have a crystalline structure consisting of four interpenetrating face-centered cubic lattices, where three of these sub lattices are filled with atoms and one with vacancies. The hH structure can be seen as a variation of the Heusler or full-Heusler lattice, where all sub lattices are filled with atoms. An example of a hH can be seen in Fig. \ref{fig:hh}.

The chemical structure of hH follows the XYZ type where X is a more electropositive transition metal, Y is a less electropositive transition metal and Z is a main group element \cite{Bos_2014}. On the other hand, full-Heuslers have a similar structure of the XY$_2$Z type.

% Adicionar código de cores na legenda.

%As previously mentioned, this work was carried out under the DFT and DFPT formalisms implemented in the Quantum Espresso package. A 3 x 3 x 3 supercell was used, resulting in a crystal of 81 atoms for the studied materials. Both, the supercells of the defect-free material, and the defective supercell were relaxed in order to obtain their stable form. The pseudopotential applied was the Optimized Norm-Conserving Vanderbilt (ONCV) with Generalized Gradient Approximation - Perdew Burke Ernzerhof (GGA-PBE) functionals, available in the sg15 package \cite{PhysRevB.88.085117}. The convergence criteria for the kinetic energy was 130 Ry, and for the electronic part was $1.0 \times 10^{-10}$ eV, while the required cell pressure was below 0.1 kbar. Finally, the defect concentrations tested were $x \approx 0.012$ for both substitutional and interstitial defects, corresponding to 1 defect per the supercell.

\begin{figure} [!h]
	\centering
	\includegraphics[scale=0.25]{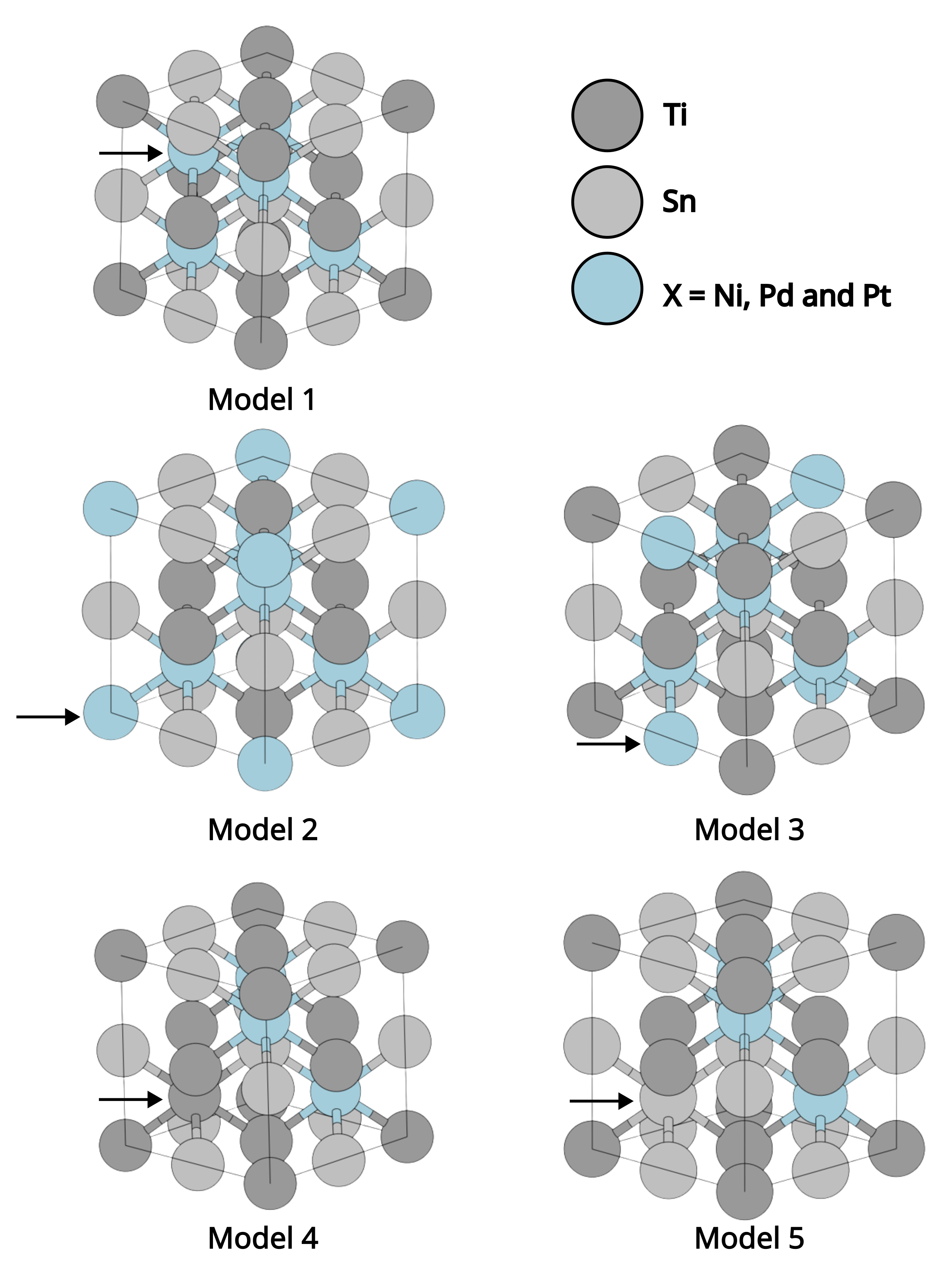}
	\caption{Defect models for TiXSn (X = Ni, Pd and Pd).}
	\label{fig:defect_models}
\end{figure}

In summary, we studied five kinds of defects for the three compounds analysed. For simplicity, these compounds are represented by TiXSn, where X can be Ni, Pd or Pt. Model 1 corresponds to an X intestitial defect in all materials. As for the substitutional defects, the model 2 represents the replacement of Ti by X. Simillarly, the model 3 considers X relpacing Sn. Moreover, model 4 consists in a Ti replacing X and, likewise, the model 5 reside in a Sn replacing X. An illustration of the models can be seen in Fig. \ref{fig:defect_models}.

\section{Methodology}\label{sec:methodology}
The DFT approach used here is based on the ansatz proposed by Kohn and Sham \cite{khonsham65}, which enabled the use of the original Hohenberg-Kohn formalism \cite{hohenberg64}. 
%which, although powerful, was not applicable due to the many body problem. The solution, proposed by Kohn and Sham was to split the many body problem into a single particle problem plus an auxiliary system, that turned DFT into an applicable and reliable method.

%DFT makes use of trial functionals in order to obtain an approximate ground state density, which can be used for properties calculation. The fact that most of the system's properties can be obtained from the ground state density is one of the strongest features of DFT.

%The self-consistent DFT method works as follows, first we provide a trial density, then the Kohn-Sham equations are solved. With the new wave function obtained, the density is recalculated and a self-consistency test is performed, if it passes then the cycle ends and the density is used to calculate the properties. Hence, this is the approximation for the density of the ground state, if the condition is not satisfied then the density becomes the new trial and the cycle repeats until self-consistency is achieved.
From the ground state density, we are able to obtain features such as the relaxed crystalline structure, where the lattice vectors and the ionic positions are optimized, as well as the electronic band structure and the density of states.

DFPT is also a self-consistent process, but in this case it is used to obtain a variation of the potential or of the density. It is used, in the present context, to obtain the phonon dispersion bands of the system, once that phonons are obtained from the variation of the Kohn-Sham potential $V_{KS}$ with respect to the ionic displacement.

Regarding the electronic structure, when we have a defect-free crystal, the band structure presents the energy bands corresponding to the periodic overlap of the energy levels of the atoms. When we introduce defects in semiconductor materials, defect states are created, which will usually be located within the gap and can be seen as a shift in the Fermi level of the material.

%As previously mentioned, this work was carried out under the DFT and DFPT formalisms implemented in the Quantum Espresso package. 
A 3 x 3 x 3 supercell was used, resulting in a crystal of 81 atoms for the studied materials. Both, the supercells of the defect-free material, and the defective supercell were relaxed in order to obtain their stable form. The pseudopotential employed was the Optimized Norm-Conserving Vanderbilt (ONCV) with Generalized Gradient Approximation - Perdew Burke Ernzerhof (GGA-PBE) functionals, available in the sg15 package \cite{PhysRevB.88.085117}. The convergence criteria for the kinetic energy was 130 Ry, and for the electronic part was $1.0 \times 10^{-10}$ eV, while the required cell pressure was below 0.1 kbar. Finally, the defect concentrations tested were $x \approx 0.012$ for both substitutional and interstitial defects, corresponding to 1 defect per the supercell.
%In addition to the effects on the electronic structure, the introduction of defects has a direct influence on transport properties, an example of which is charged defect scattering. This scattering is an important mechanism and it is essential to consider it in transport calculations that do not rely on the constant relaxation time approximation.

Several forms of defects can occur in a crystal, from which two distinct forms will be addressed, namely: interstitial and substitutional defects. Interstitial defects concern the appearance of atoms on sites that do not correspond to the patterns of the underlying crystal structure. In particular, in hH materials interstitial defects can arise in Wyckoff positions corresponding to vacancies that are occupied in Heusler materials. On the other hand, substitutional defects correspond to the exchange of an atom on a certain site by an atom of another species of the crystal. In our case, the substitutions performed were X replacing Ti and Sn in TiXSn crystals where X = Ni, Pd and Pt. We also tested the substitution of Ti and Sn replacing X. It is worth noting that the defect atom does not necessarily have to be an atom present on the existent chemical composition of the crystal, but this case is not dealt in this work.

Finally, in order to estimate whether the defects are favorable to arise or not, we must be able to estimate the formation energy of these defects in the crystal. The expression for the formation energy of interstitial defects, is given by \cite{Rittiruam_Yangthaisong_Seetawan_2018, Dey21},
\begin{equation}
	\label{eq:inter}
	\Delta E_{I} = E_{def} - [E_{pure} + \sum_x E_{bulk}(x)],
\end{equation}
where $E_{def}$ is the total energy of the defective crystal, $E_{pure}$ is the total energy for the defect-free crystal and $E_{bulk}$ is the total energy of the bulk of the defect, where the sum is for the number of atoms of the $x$ atomic species in the interstitial site. On the other hand, for substitutional defects we can calculate the formation energy as follows,
\begin{equation}
	\label{eq:subs}
	\Delta E_{S} = E_{def} + \sum_{x_1} E^{S}_{bulk}(x_1) - [E_{pure} + \sum_{x_2} E^{def}_{bulk}(x_2)],
\end{equation}
where $E^{S}_{bulk}$ is the total energy of a bulk of the replaced atom and $x_1$ is the species of the replaced atom, $E^{def}_{bulk}$ is the total energy of a defect atom bulk and $x_2$ is the species of the defect atom, the other variables follow the previous notation.

\section{Results and Discussion}\label{sec:results}

First, we discuss results related to the structure of the material. Using the `vc-relax' method available in QE, the relaxation of the unit cell is carried out by changing the lattice vectors and ionic positions. Initially, we performed a relaxation of the defect-free supercell of each one of the three materials. The result obtained in this first relaxation was then used as the initial structure for the defective supercells. The initial and final values of the lattice parameters with the corresponding variations are presented in Table \ref{tab:lparam}. It should be noted that these values correspond to 3 x 3 x 3 supercells obtained by scaling primitive unit cells.

\begin{table}[!h] 
	\caption{Initial and final lattice parameters obtained from the relaxation process and the corresponding changes in the lattice parameter ($\Delta a/a$).}
	\begin{tabular}{rccc}
		\multicolumn{1}{c}{Model}      & $a_0$ (Å) & $a_f$ (Å) & \multicolumn{1}{r}{$\Delta a/a (\%)$} \\ \hline
		\multicolumn{1}{c}{TiNiSn (1)} & 12.530    & 12.646    & 0.00926                               \\
		(2)                            &           & 12.627    & 0.00774                               \\
		(3)                            &           & 12.610    & 0.00638                               \\
		(4)                            &           & 12.678    & 0.01181                               \\
		(5)                            &           & 12.696    & 0.01325                               \\ \hline
		\multicolumn{1}{c}{TiPdSn (1)} & 13.182    & 13.206    & 0.00182                               \\
		(2)                            &           & 13.190    & 0.00061                               \\
		(3)                            &           & 13.174    & -0.00061                              \\
		(4)                            &           & 13.206    & 0.00182                               \\
		(5)                            &           & 13.228    & 0.00349                               \\ \hline
		\multicolumn{1}{c}{TiPtSn (1)} & 13.222    & 13.251    & 0.00219                               \\
		(2)                            &           & 13.230    & 0.00061                               \\
		(3)                            &           & 13.215    & -0.00053                              \\
		(4)                            &           & 13.246    & 0.00182                               \\
		(5)                            &           & 13.267    & 0.00340                               \\ \hline
	\end{tabular}
	\label{tab:lparam}
\end{table}

\begin{figure*}[!h]
	\centering
	\includegraphics[scale=0.19]{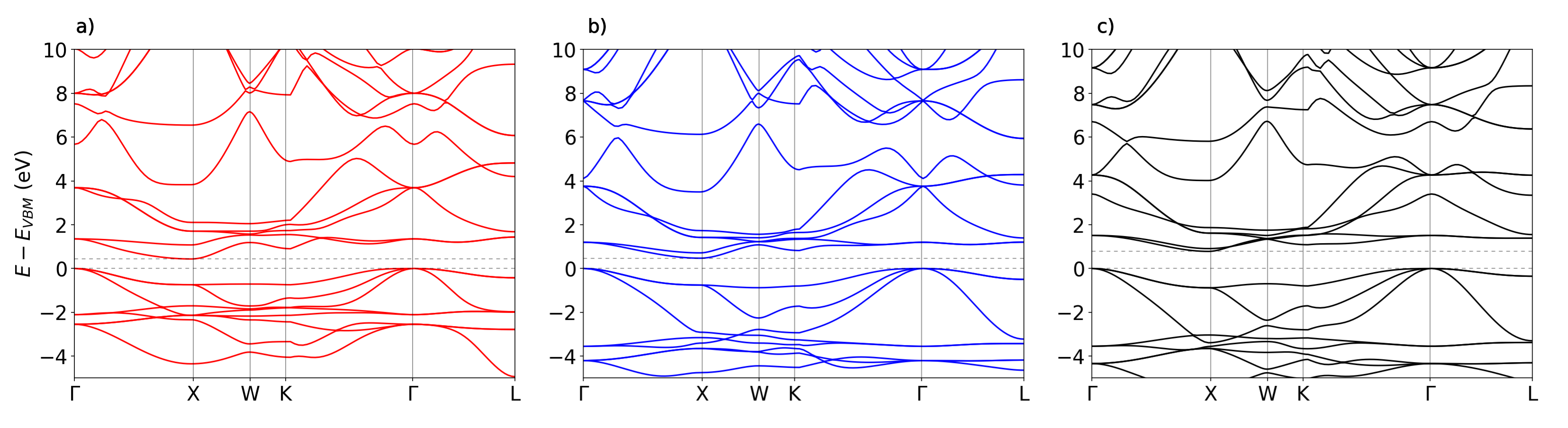}
	\caption{Electronic Band Structure of a)TiNiSn b)TiPdSn c)TiPtSn. The dotted lines indicate de band gap.}
	\label{fig:electronic}
\end{figure*}

\begin{figure} [!h]
	\centering
	\includegraphics[scale=0.40]{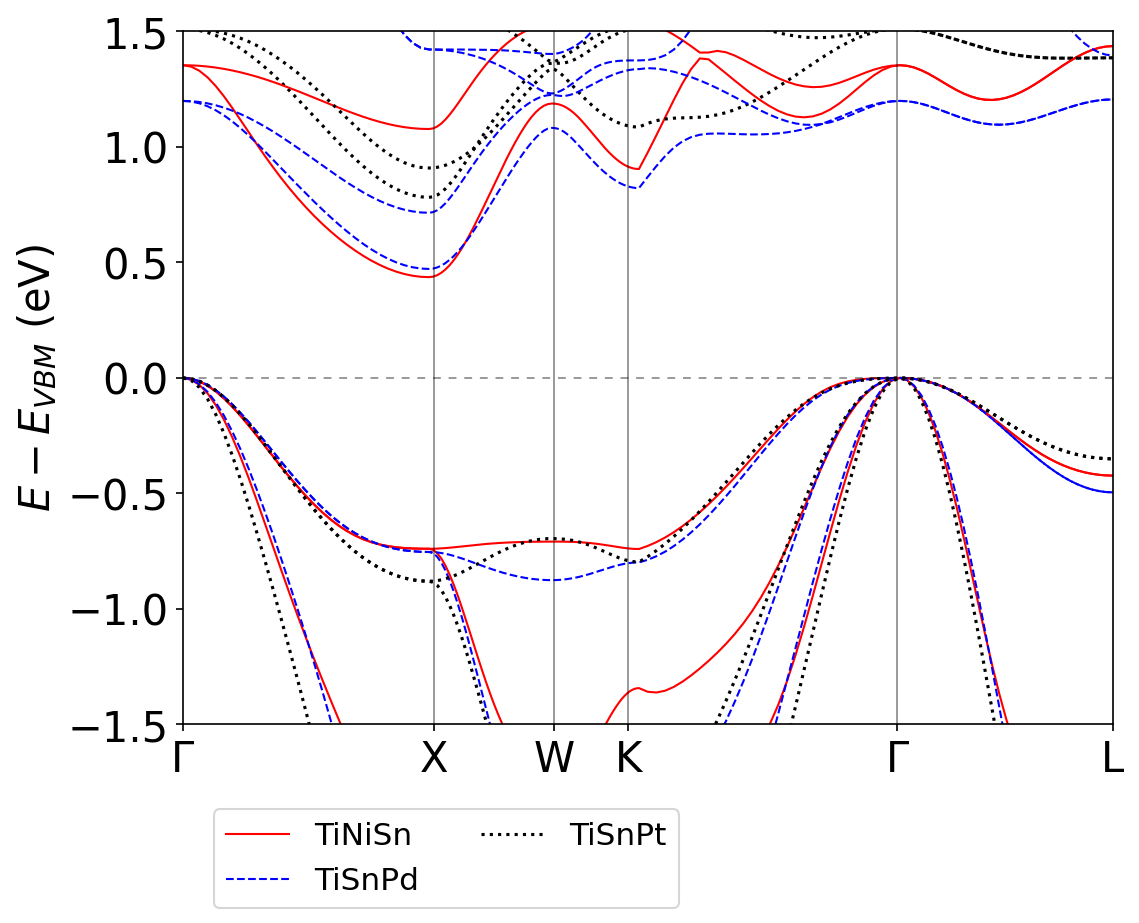}
	\caption{Comparison of electronic band structures for TiXSn materials where X = Ni, Pd and Pt.}
	\label{fig:comparsion_bands}
\end{figure}

The electronic band structure, near the band edges, obtained using GGA-PBE exchange-correlation functional, through the aforementioned DFT methodology, for TiNiSn, TiPdSn and TiPtSn can be seen separately in Fig. \ref{fig:electronic}, and superposed in the Fig. \ref{fig:comparsion_bands}. The three materials are indirect gap semiconductors with the valence band maximum (VBM) at $\Gamma$ and the conduction band minimum (CBM) at $X$. The calculated gaps for the materials are $0.44$ eV, $0.47$ eV and $0.78$ eV, for TiNiSn, TiPdSn and TiPtSn, respectively.

It is well known that DFT using functionals such as LDA, PBE or PBEsol tends to underestimate the material gap \cite{perdew85}, therefore we also performed calculations using hybrid functionals such as HSE06 (Heyd-Scuseria-Ernzerhof) \cite{heyd03}, available in the QE package. The results obtained for the gap were $0.53$ eV, $0.70$ eV and $1.28$ eV, for TiNiSn, TiPdSn and TiPtSn, respectively. They are in good agreement with results obtained previously \cite{Dasmahapatra20} of $0.76$ eV and $1.13$ eV for TiPdSn and TiPtSn respectively.

By examining the top of the valence bands from the electronic band structures shown in Fig. \ref{fig:comparsion_bands}, we can see a great similarity between the materials. One can see the merging of two bands, which in terms of effective mass can represent the heavy hole, related to the band with the widest concavity, while the light hole corresponds to the band with the sharpest concavity. It is also worth noting that TiPtSn has a smaller light hole effective mass. 

Inspecting the minima of the conduction bands, we also see a great similarity between the materials, indicating similar behavior of the electrons in the different systems, since their effective masses are similar.

In both cases, we can see relatively wide bands that indicate low effective masses, implying reduced carrier mobility, which is not ideal when looking for high conductivities. This type of property can be improved through processes such as doping or nanostructuring. 

\begin{figure*}[!h]
	\centering
	\includegraphics[scale=0.19]{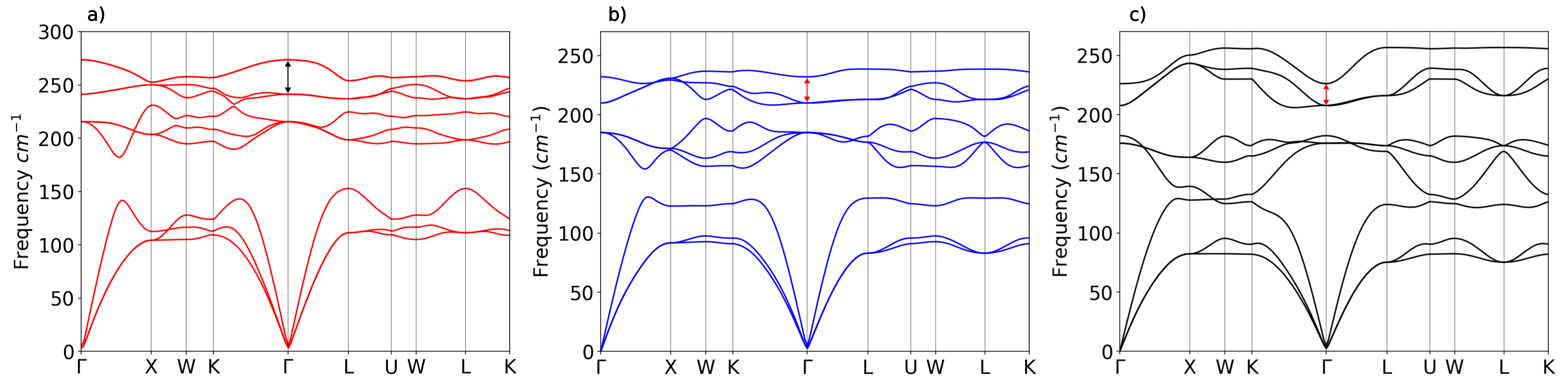}
	\caption{Phonon dispersion of a)TiNiSn b)TiPdSn c)TiPtSn. The double arrows indicate the LO-TO splitting. }
	\label{fig:phonons}
\end{figure*}

Regarding phonon dispersion, we can see the results in Fig. \ref{fig:phonons} for TiXSn, where X = Ni, Pt and Pd. In all cases, the LO-TO splitting correction has already been taken into account, which has led to the degeneracy breaking of the higher frequency optical modes.

%The LO-TO splitting correction is given by the LST (Lyddane-Sachs-Teller) equation,
%\begin{equation}
%    \frac{\omega_{LO}^2}{\omega_{TO}^{2}} = \frac{\epsilon_0}{\epsilon_\infty},
%\end{equation}
%which gives the relationship between the longitudinal optical frequency $\omega_{LO}$ and the transverse optical frequency $\omega_{TO}$ on the left-hand side, in terms of the ratio of the static permittivity $\epsilon_0$ to the permittivity $\epsilon_{\infty}$ when $\omega \rightarrow \infty$.

The LO-TO splitting correction is given by a non-analytical
term correction \cite{Gonze_Lee_1997}, which depends on the Born effective charges $Z_{\alpha\beta}$
and the dielectric constant $\epsilon_{\alpha\beta}$. The fomulation is given by,

\begin{equation}
	C^{NA}_{\kappa \alpha, \kappa' \beta}(\vec{q} \rightarrow 0) = \frac{4\pi}{\Omega_0} \frac{(\sum_{\gamma} q_\gamma Z^{}_{\kappa, \gamma \alpha})(\sum_{\gamma'} q_{\gamma'} Z^{*}_{\kappa', \gamma' \beta})}{\sum_{\alpha \beta} q_{\alpha} \epsilon^{\infty}_{\alpha \beta} q_{\beta} },
\end{equation}
where $q_{i}$ is the components of the phonon eigenvector, $Z^{*}_{\kappa, \gamma \alpha}$ is the Born effective charge tensor elements, $\epsilon^{\infty}_{\alpha \beta}$ is the dielectric tensor elements and $\Omega_0$ is the Brillouin-Zone volume.

By examining the separation between the highest frequency transverse and longitudinal optical modes of the materials, one can see that the LO-TO splitting effect decreases according to the order: TiNiSn $\rightarrow$ TiPdSn $\rightarrow$ TiPtSn. This is because LO-TO splitting is caused by the long-range nature of the Coulomb dipole interaction, mostly present in LO modes, which is proportional to the degree of ionicity of the compound. In other words, this is consistent with the decreasing Pauling electronegativity of the elements: Pt (2.28), Pd (2.20), and Ni (1.91) \cite{Huheey1993}. The numerical values of the LO-TO splitting for TiNiSn, TiPdSn, and TiPtSn are $31.86$ cm\textsuperscript{-1}, $22.41$ cm\textsuperscript{-1}, and $18.42$ cm\textsuperscript{-1}, respectivelly.

As previously mentioned, the introduction of impurities into the material can be a determining factor in understanding and even improving its transport properties. These defects can change the electronic structure and the dispersion of phonons previously presented, as well as increasing the impurity scattering rate, therefore, it is important to understand the occurrence of defects in these materials.

\begin{figure} [!h]
	\centering
	\includegraphics[scale=0.43]{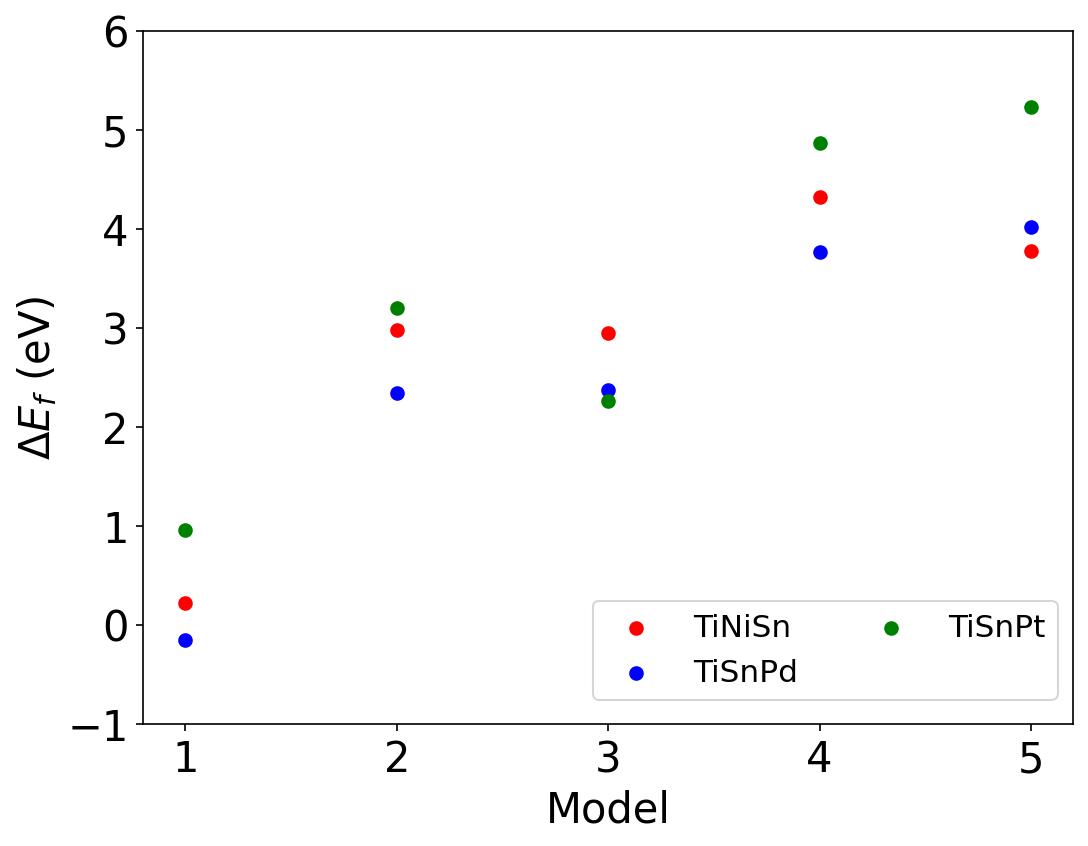}
	\caption{Formation energies as functions of the models for TiNiSn, TiPdSn and TiPtSn. Model 1 corresponds to interstitial defects, models 2, 3, 4 and 5 correspond to substitutional models as explained in section \ref{sec:materials}.}
	\label{fig:formation_results}
\end{figure}

The results for the formation energy calculations using the total energies from DFT calculations and equations (\ref{eq:inter}) and (\ref{eq:subs}) are presented in Fig. \ref{fig:formation_results}. The formation energies of the Pd interstitial defects in TiPdSn and Ni interstitial defects TiNiSn are very low and even negative in the case of TiPdSn.

Regarding the defect formation energies in TiNiSn, we would like to point out that the formation energy of Ni interstitial defect being $\Delta E_f \approx 0.21$ eV is in good agreement with results previously obtained by Hazama \textit{et al.}\cite{Hazama_Asahi_Matsubara_Takeuchi_2010}, Douglas \textit{et al.}\cite{douglas14}, Dey and Dutta \cite{Dey21}, which validates the methodology used.

 The formation energy for interstitial Pd defects in TiPdSn, with a value of $\Delta E_f \approx -0.14$ eV, stands out. This unexpected negative value indicates a high propensity to form these defects in the material. This finding suggests that, since the interstitial defects occupy sites that are occupied in full-Heusler compounds, hH TiPdSn may be difficult to be synthesized. In other words, although hH TiPdSn is dynamically stable, which was verified by the absence of negative frequencies in the phonon modes, it may not be thermodinamically stable with respect to the full-Heusler compound. 
 
 On the other hand, the formation energy  of Pt interstitial defects in TiPtSn is much higher, with a value of $\Delta E_f \approx 0.91$ eV, in contrast to the other two compounds. This suggests that interstitial defects in TiPtSn, due to their low concentration, may not have a relevant role in determining the band gap of the material, as it happens in the case of TiNiSn. 
 
 The substitutional defects have, in all cases, much higher formation energies, ranging from $\approx 2.0$ eV up to $\gtrsim 5.0$ eV, and will occur in much lower concentrations.
 
\begin{figure} [!h]
 	\centering
 	\includegraphics[scale=0.40]{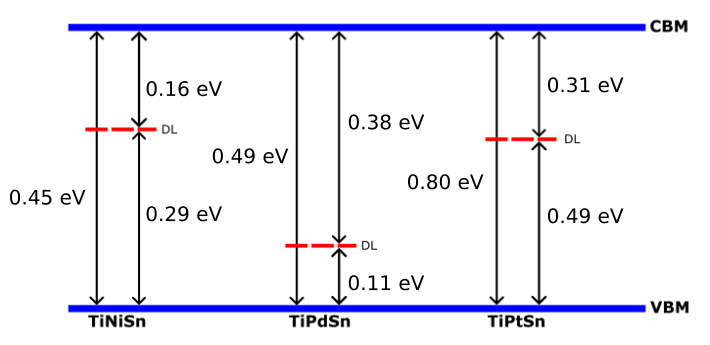}
 	\caption{Intestital defect energy level position relative to band edges for TiNiSn, TiPdSn and TiPtSn.}
 	\label{fig:defect_level}
 \end{figure}

 Fig. \ref{fig:defect_level} shows schematically our results for the band edges, conduction band minumum (CBM) and valence band maximum (VBM), together with de defect level (DL) introduced in the band gap by interstitials obtained using the GGA-PBE functional for the three materials. As mentioned before, these results for the band gap are smaller than those calculated using the HSE06 functional. In the case of TiNiSn, our calculated band gap, $0.45$ eV, and the DL position relative to CBM, $0.16$ eV, are similar to the results obtained by Dey and  Dutta \cite{Dey21}, $0.43$ eV and $0.10$ eV, respectively. Therefore, according to our calculations, the band gap for this compound, assuming its change due to the presence of interstials in a high concentration, would be of $0.16$ eV, which is close to the experimental value of $0.12$ eV \cite{Aliev1990}.
 
 For the TiPdSn compound, the situation is quite distinct, the DL position is close to the VBM, suggesting that this level should be similar to a bulk level. This is consistent with the picture that the full-Heusler TiPdSn compound should be more stable than the hH one. 
 
 In the case of TiPtSn, the DL is closer to the CBM, suggesting a reduction in the band gap from $0.80$ eV to $0.31$ eV, however, since the interstitial formation energy in this case is relatively high, a reduced band gap in the experiment should not be expected, in contrast to the case of TiNiSn.
 
\section{Conclusions}\label{sec:conclusions}
%Looking at the results of the electronic band structure, we could see that the materials have very similar effective masses of holes, except for TiPtSn, which has a slightly lower mass than the others. In addition, we could see that the effective masses of the electrons are similar, which means that the electrons will behave similarly in both materials.

Electronic, vibrational, and defect properties of the hH compounds TiXSn (X = Pd, Pt and Ni) were investigated using first-principles total energy calculations. The three compounds have similar band structures, which are agreement with previous calculations \cite{Dey21,Xiong2022,Xiong2024}. 
%The interstitial defect level in TiNiSn also agrees with previous results \cite{Dey21}.

%Regarding phonon scattering, we conclude that, based on the difference in the LO-TO splitting correction, the TiPtSn probably has the highest concentration of carriers among the materials. This is due to the fact that the correction in this material is lower than in the others, indicating that the screening effect is greater.

Our calculations of the phonon dispersion included the so-called LO-TO splitting, which affects the optical phonon bands. The acoustic phonon bands are in agreement with previous results in the literature \cite{Xiong2024}. Our results show that the LO-TO splitting decreases according to the order: TiNiSn $\rightarrow$ TiPdSn $\rightarrow$ TiPtSn, which follows the increasing order in the Pauling electronegativity of the elements: Ni, Pd, and Pt. Our calculations indicate that TiNiSn has greater ionic character, therefore, carrier scattering by polar optical phonons in TiNiSn should be more relevant than in the other compounds. 

Our results for native defects show interesting differences between the three compounds. In the case of TiNiSn, the Ni interstitial has a very small formation energy of $0.21$ eV in agreement with previous calculations \cite{Hazama_Asahi_Matsubara_Takeuchi_2010,douglas14,Dey21}, indicating a high concentration of these defects that affects the electronic structure of the material. Surprisingly, the formation energy of the Pd interstitial in TiPdSn is negative $-0.14$ eV, strongly suggesting that synthesis of the material in its hH structure should be difficult. On the other hand, the formation energy of the Pt interstitial in TiPtSn is higher $0.91$ eV, indicating that these defects should not have a strong influence in the electronic structure of the material. Substitutional defects have much higher formation energies and should appear in very low concentrations. They tend to be irrelevant for the synthesis, electronic structure, or carrier transport properties of these materials.

%Based on the formation energies obtained, it was possible to see that materials such as TiPdSn and TiNiSn are unlikely to be defect-free synthesized, since both have point defects with very low formation energies. On the other hand, TiPtSn defects has a substantially higher formation energy, even in terms of interstitial defects.

%We have also added to the bibliography electronic structure and phonon dispersion results for TiXSn where X = Pd and Pt, these materials having few published results to the authors' knowledge. 

The results obtained, in addition to improving the basic knowledge about these materials, can also provide important information for experimentalists in future attempts to synthesize them. %, both with regard to vibrational and electronic properties, and what to expect with regard to the formation of defects in the process.

%\section{CRediT authorship contribution statement}\label{credit_authorship}

%\section{Declaration of competing interest}\label{competing_interest}
%The authors declare that they have no known competing financial interests or personal relationships that could have appeared to influence the work reported in this paper.

%\section{Data availability}\label{data_availability}
%Data will be made available on request.

\section{Acknowledgement}\label{sec:acknowledgement}
AA gratefully acknowledges support from the Brazilian agencies CNPq and FAPESP under Grants \#2010/16970-0, \#2013/08293-7, \#2015/26434-2, \#2016/23891-6, \#2017/26105-4,and \#2019/26088-8. MCL gratefully acknowledges support from the Brazilian agency CAPES. The calculations were performed at CCJDR-IFGW-UNICAMP in Brazil.
%We would like to thank CAPES, FAPESP and CNPq for the research funds, the IFGW for the support and the CCJDR-IFGW in Campinas where the calculations were performed.


%%       \citet{<label>} ==> Jones et al. [21]
%%       \citep{<label>} ==> [21]
%%

\bibliographystyle{elsarticle-num-names} 
\bibliography{Manuscript_TiXSn_JPCS.bib}

%\begin{thebibliography}{00}



%\end{thebibliography}

\end{document}